\documentclass[prl,a4paper,twocolumn,showpacs]{revtex4}%
\usepackage{amsmath}
\usepackage{graphicx}
\usepackage{dcolumn}
\usepackage{bm}
\usepackage{amsfonts}
\usepackage{amssymb}%

\begin{document}
\title{Voltage generation by ferromagnetic resonance}
\author{Xuhui Wang}
\author{Gerrit E. W. Bauer}
\affiliation{Kavli Institute of NanoScience, Delft University of Technology, 2628 CJ Delft,
The Netherlands}
\author{Bart J. van Wees}
\affiliation{Department of Applied Physics and Materials Science Center, University of
Groningen, Nijenborgh 4, 9747 AG Groningen, The Netherlands}
\author{Arne Brataas}
\affiliation{Department of Physics, Norwegian University of Science and Technology, N-7491
Trondheim, Norway}
\author{Yaroslav Tserkovnyak}
\affiliation{Lyman Laboratory of Physics, Harvard University, Cambridge, MA 02138, USA}
\date{\today}

\begin{abstract}
A ferromagnet can resonantly absorbs rf radiation to sustain a steady
precession of the magnetization around an internal or applied magnetic field.
We show that under these ferromagnetic resonance (FMR) conditions, a dc
voltage is generated at a normal-metal electric contact to a ferromagnet with
spin-flip scattering. This mechanism allows an easy electric detection of
magnetization dyamics.

\end{abstract}

\pacs{76.50.+g, 72.25.Mk, 73.23.-b, 73.40.-c}
\maketitle

The field of magnetoelectronics utilizes the electronic spin degrees of
freedom to achieve new functionalities in circuits and devices made from
ferromagnetic and normal conductors. The modulation of the DC electrical
resistance by means of the relative orientation of the magnetizations of
individual ferromagnetic elements (\textquotedblleft giant
magnetoresistance\textquotedblright) is by now well-established. Dynamic
effects, such as the current-induced magnetization reversal, are still subject
of cutting edge research activities. Here we concentrate on an application of
the concept of spin-pumping, \textit{i.e}. the emission of a spin current from
a moving magnetization of a ferromagnet (F) in electrical contact with a
normal conductor (N) \cite{tserkovnyak-pumping, tserkovnyak-pumping-rmp},
\textit{viz}. the \textquotedblleft spin battery\textquotedblright%
\  \cite{brataas-sb}. In this device a ferromagnet that precesses under
ferromagnetic resonance (FMR) conditions pumps a spin current into an attached
normal metal that may serve as a source of a constant spin accumulation (see
also Ref. \onlinecite{Watts2005}). In this Letter we report that spin-flip
scattering in the ferromagnet translates the pumped spin accumulation into a
charge voltage over an F$|$N junction. Due to the spin-flip scattering in F, a
back-flow spin current collinear to the magnetization is partially absorbed in
the ferromagnet. Since the interface and bulk conductances are spin-dependent,
this leads to a net charging of the ferromagnet, which thus serves as a source
as well as electric analyzer of the spin pumping current. We note the analogy
to the voltage in excited F$|$N$|$F spin valves predicted by Berger
\cite{berger1999} and recently analyzed by Kupferschmidt \textit{et al.}
\cite{brouwer}. Since the spin-flip scattering in conventional magnets such as
permalloy is very strong, this effect provides a handle to experimentally
identify the FMR induced spin accumulation in the simplest setup
\cite{Azevedo2005}. A detailed experimental test of our predictions is in
progress \cite{Costache}.

\begin{figure}[ptb]
\includegraphics[scale=0.45]{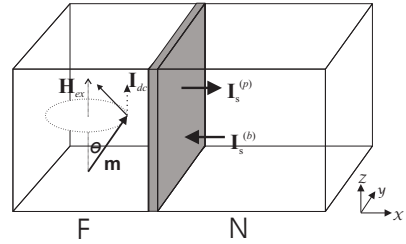} \caption{Schematic view of spin
battery operated by ferromagnetic resonance. The dotted line $\mathbf{I}_{dc}$
represents the dc component of pumping current. }%
\label{fig:geo}%
\end{figure}

The \textquotedblleft spin battery\textquotedblright~operated by ferromagnetic
resonance has been proposed by Brataas \textit{et al}. \cite{brataas-sb} in
the limit of weak spin flip scattering in the ferromagnet. It is based on the
spin current pumped into a normal metal by a moving magnetization
($\text{F}|\text{N}$) \cite{tserkovnyak-pumping}
\begin{equation}
\mathbf{I}_{s}^{(p)}=\frac{\hbar}{4\pi}\left(  \operatorname{Re}%
g^{\uparrow \downarrow}\mathbf{m}\times \frac{d\mathbf{m}}{dt}+\operatorname{Im}%
g^{\uparrow \downarrow}\frac{d\mathbf{m}}{dt}\right)  ~,
\end{equation}
where $\mathbf{m}$ is the unit vector of magnetization. $\operatorname{Re}%
g^{\uparrow \downarrow}$ and $\operatorname{Im}g^{\uparrow \downarrow}$ are the
real and imaginary parts of the (dimensionless) spin-mixing conductance
$g^{\uparrow \downarrow}$ \cite{brataas-ct}. This spin current creates a spin
accumulation $\mathbf{s}$ in the normal metal, which induces a backflow of
spins, and, as we will see, charges the ferromagnet. According to
magnetoelectronic circuit theory \cite{brataas-ct} the charge and spin
currents flowing through the $\text{F}|\text{N}$ interface (into N) in the
presence of non-equilibrium charge and spin accumulations $\mu_{0}^{N},$
$\mathbf{s}$ in N and $\mu_{0}^{F}$, $\mu_{s}^{F}\mathbf{m}$ in F, read
\cite{brataas-ct}
\begin{align}
I_{c}  &  =\frac{eg}{2h}\left[  2(\mu_{0}^{F}-\mu_{0}^{N})+p\mu_{s}%
^{F}-p(\mathbf{m}\cdot \mathbf{s})\right] \nonumber \\
\mathbf{I}_{s}^{(b)}  &  =\frac{g}{8\pi}\left[  2p(\mu_{0}^{F}-\mu_{0}%
^{N})+\mu_{s}^{F}-(1-2\operatorname{Re}g^{\uparrow \downarrow}/g)\mathbf{m}%
\cdot \mathbf{s}\right]  \mathbf{m}\nonumber \\
&  -\frac{\operatorname{Re}g^{\uparrow \downarrow}}{4\pi}\mathbf{s}%
-\frac{\operatorname{Im}g^{\uparrow \downarrow}}{4\pi}(\mathbf{s}%
\times \mathbf{m})~,
\end{align}
where $g=g^{\uparrow}+g^{\downarrow}$ is the total interface conductance of
spin-up and spin-down electrons, $p$ is the contact polarization given by
$p=(g^{\uparrow}-g^{\downarrow})/(g^{\uparrow}+g^{\downarrow})$. For typical
metallic interfaces, the imaginary part of the mixing conductance is quite
small \cite{kelly}, hence discarded in the following discussion . We choose
the transport direction along the $x$-axis that is perpendicular to the
interface at the origin. $\mathbf{H}_{ex}$, the sum of DC external and
uniaxial anisotropy magnetic fields, points in the $z$-direction, which is
also the chosen spin quantization axis in the normal metal. At the
ferromagnetic resonance, the magnetization precesses steadily around the
$z$-axis with azimuthal angle $\theta$ (see Fig. \ref{fig:geo}) that is
tunable by the intensity of an AC magnetic field. The thickness of the normal
and ferromagnetic metal films are $d_{N}$ and $d_{F}$, respectively.
$\mathbf{s}\left(  x,t\right)  $ is determined by the spin-diffusion equation
\cite{johnson-silsbee}
\begin{equation}
\frac{\partial \mathbf{s}}{\partial t}=D_{N}\frac{\partial^{2}\mathbf{s}%
}{\partial x^{2}}-\frac{\mathbf{s}}{\tau_{sf}^{N}}~, \label{eq:spindiffusion}%
\end{equation}
where $\tau_{sf}^{N}$ is the spin-flip relaxation time and $D_{N}$ the
diffusion constant in the normal metal. Assuming that the magnetization
precesses around the $z$-axis with angular velocity $\omega$, we consider the
limit where the spin-diffusion length in the normal metal is much larger than
the transverse spin-averaging length $l_{\omega}\equiv \sqrt{D_{N}/\omega}$,
\textit{i.e}., $\lambda_{sd}^{N}\gg l_{\omega}$, or equivalently $\omega
\tau_{sf}^{N}\gg1$. We can then distinguish two regimes. When the thickness of
the normal metal $d_{N}\gg l_{\omega}$, which is equivalent to the Thouless
energy $\hbar D_{N}/d_{N}^{2}\ll \hbar \omega$, the oscillating transverse
component of the induced spin accumulation vanishes inside the normal metal,
and one is left with a time-dependent spin accumulation along $z$-axis
decaying away from the interface on the scale $\lambda_{sd}^{N}$. The backflow
due to the steady state spin accumulation aligned along the $z$-axis cancels
the same component of the pumping current. The former acquires the universal
value $\hbar \omega$ when the spin-flip scattering is sufficiently weak
\cite{brataas-sb}. The opposite regime of ultrathin or ultraclean normal metal
films in which $\hbar D_{N}/d_{N}^{2}\gg \hbar \omega$ the spin accumulation
$\mathbf{s}$ is governed by a Bloch equation and will be discussed elsewhere
\cite{wangbloch}.

Continuity of the total spin current into the normal metal at the interface
\begin{equation}
\mathbf{I}_{s}=\mathbf{I}_{s}^{(p)}+\mathbf{I}_{s}^{(b)}
\label{eq:totalcurrent}%
\end{equation}
is the first boundary condition for the diffusion equation, ${\partial
\mathbf{s}}/{\partial x}|_{x=0}=-2\mathbf{I}_{s}/(\hbar \nu_{\mathrm{dos}%
}AD_{N}),$ where $\nu_{\mathrm{dos}}$ is the one-spin density of states and
$A$ the area of the interface, and the second is its vanishing at the sample
edge ${\partial \mathbf{s}}/{\partial x}|_{x=d_{N}}=0$. The time-averaged
solution of Eq. (\ref{eq:spindiffusion}) reads $\left \langle {\mathbf{s}%
}\right \rangle _{t}=s_{z}\mathbf{\hat{z}}$ with
\begin{equation}
s_{z}=\frac{\cosh \left(  x-d_{N}\right)  /\lambda_{sd}^{N}}{\sinh
d_{N}/\lambda_{sd}^{N}}\frac{2\lambda_{sd}^{N}}{\hbar \nu_{\mathrm{dos}}AD_{N}%
}I_{s,z}~.
\end{equation}

\begin{figure}[ptb]
\includegraphics[scale=1.0]{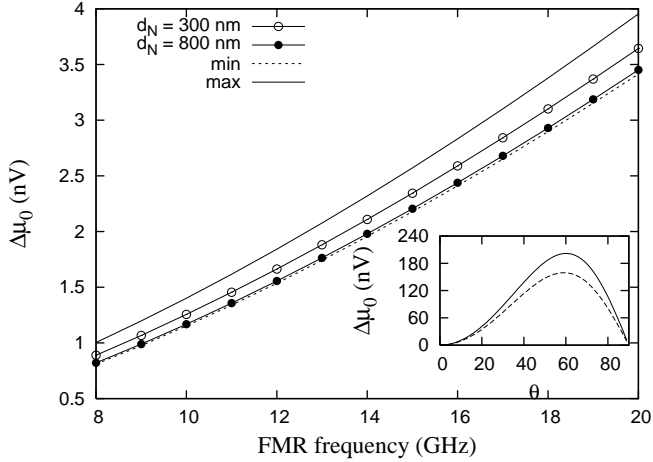}\caption{The voltage
drop (in $\operatorname{nV}$) as function of FMR frequency (in
$\operatorname{GHz}$) for Py$|$Al interface. The line with circles denotes the
situations when $d_{N}=300~\operatorname{nm}$ (empty symbols) and
$d_{N}=800~\operatorname{nm}\text{ }$(filled symbols) when the thickness of
ferromagnet is taken as $d_{F}=14~\operatorname{nm}$. The solid and dashed
lines refer to the limits as indicated by Eq. (\ref{eq:volmax}) and Eq.
(\ref{eq:volmin}). These curves indicate that due to averaging of the
transverse spin components inside the normal metal, the voltage is not linear
with FMR frequency. The precession angle of magnetization is taken as
$\theta=5^{\circ}$. The inset shows the angle dependence of the voltage at
fixed frequency $15.5\operatorname{GHz}$. At small angle, the voltage drop is
proportional to $\theta^{2}$.}%
\label{fig:pyalvol-omega}%
\end{figure}

\begin{figure}[ptb]
\includegraphics[scale=1.0]{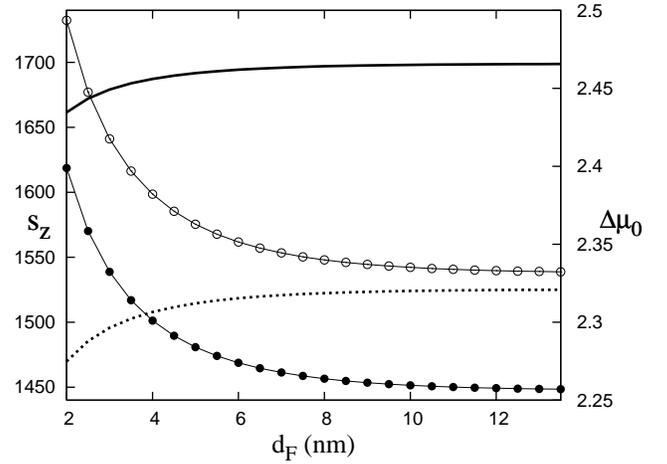} \caption{Lines with circles are
the spin-pumping induced accumulation $s_{z}/e$ (in unit of $\operatorname{nV}%
$) in Al near the interface to permalloy as a function of the Py layer
thickness $d_{F}$ and two Al layer thicknesses, \textit{i.e.}, $d_{N}%
=300\operatorname{nm}$ (empty symbols) and $d_{N}=800\operatorname{nm}$
(filled symbols). Solid($d_{N}=300\operatorname{nm}$)and dotted($d_{N}%
=800\operatorname{nm}$) lines are the chemical potential discontinuity across
the interface $\Delta \mu_{0}/e$ (in units of $\operatorname{nV}$), as a
function of the Py layer thickness $d_{F}$. The FMR frequency is 15.5
$\operatorname{GHz}$.}%
\label{fig:pyal}%
\end{figure}

The component of the spin accumulation parallel to the magnetization is a
constant for the precessional motion considered here. It can penetrate the
ferromagnet, hence building up a spin accumulation $\mu_{s}^{F}=\mu_{\uparrow
}^{F}-\mu_{\downarrow}^{F}$ in F, which obeys the spin diffusion equation
\cite{johnson-silsbee}
\begin{equation}
\frac{\partial^{2}\mu_{s}^{F}(x)}{\partial x^{2}}=\frac{\mu_{s}^{F}%
(x)}{\left(  \lambda_{sd}^{F}\right)  ^{2}}~,\label{eq:sdeferro}%
\end{equation}
where $\lambda_{sd}^{F}$ is the spin-flip diffusion length in the ferromagnet.
The boundary conditions are given by the continuity of the longitudinal spin
current at the interface
\begin{equation}
\sigma_{\uparrow}\left(  \frac{\partial \mu_{\uparrow}^{F}}{\partial x}\right)
_{x=0}-\sigma_{\downarrow}\left(  \frac{\partial \mu_{\downarrow}^{F}}{\partial
x}\right)  _{x=0}=-\frac{2e^{2}}{\hbar A}I_{s,z}\cos \theta
\end{equation}
and a vanishing spin current at the outer boundary
\begin{equation}
\sigma_{\uparrow}\left(  \frac{\partial \mu_{\uparrow}^{F}}{\partial x}\right)
_{x=-d_{F}}-\sigma_{\downarrow}\left(  \frac{\partial \mu_{\downarrow}^{F}%
}{\partial x}\right)  _{x=-d_{F}}=0~,
\end{equation}
where $\sigma_{\uparrow(\downarrow)}$ is the conductivity of spin up (down)
electrons in the ferromagnet \cite{yt-stiffness}. In the steady state there
can be no net charge flow. From $I_{c}=0$ follows that a charge chemical
potential difference $\mu_{0}^{F}-\mu_{0}^{N}=p[s_{z}\cos \theta-\mu_{s}%
^{F}]_{x=0}/2$ builds up across the contact. At the interface on the F side,
the longitudinal component of the total spin current leaving the ferromagnet
then reads
\begin{equation}
I_{s,z}\cos \theta=\frac{(1-p^{2})g}{8\pi}[\mu_{s}^{F}-s_{z}\cos \theta
]_{x=0}~.\label{eq:long}%
\end{equation}
The interface resistance is in series with a resistance $\rho_{\omega
}=l_{\omega}/(h\nu_{\mathrm{dos}}AD_{N})$ of the bulk normal metal of
thickness $l_{\omega}$ that accounts for the averaging of the transverse spin
current components. This reduces the interface conductances for spin-up (down)
electrons to $g_{\omega}^{\uparrow(\downarrow)}=g^{\uparrow(\downarrow
)}/(1+\rho_{\omega}g^{\uparrow(\downarrow)})$ and the spin-mixing conductance
$g_{\omega}^{\uparrow \downarrow}=\operatorname{Re}g^{\uparrow \downarrow
}/(1+\rho_{\omega}\operatorname{Re}g^{\uparrow \downarrow}).$ We also
introduce
\begin{equation}
g_{\omega}=g_{\omega}^{\uparrow}+g_{\omega}^{\downarrow},\quad p_{\omega
}=\frac{g_{\omega}^{\uparrow}-g_{\omega}^{\downarrow}}{g_{\omega}^{\uparrow
}+g_{\omega}^{\downarrow}}~.
\end{equation}
Solving Eq. (\ref{eq:sdeferro}) under the above boundary conditions gives
\begin{equation}
\mu_{s}^{F}(x)=\frac{\tilde{g}\cosh \left[  \left(  x+d_{F}\right)
/\lambda_{sd}^{F}\right]  \cos \theta}{[\tilde{g}+g_{F}\tanh \left(
d_{F}/\lambda_{sd}^{F}\right)  ]\cosh \left(  d_{F}/\lambda_{sd}^{F}\right)
}s_{z}|_{x=0}\label{eq:saf}%
\end{equation}
where $\tilde{g}=(1-p_{\omega}^{2})g_{\omega}$ and $g_{F}=4hA\sigma_{\uparrow
}\sigma_{\downarrow}/[e^{2}\lambda_{sd}^{F}(\sigma_{\uparrow}+\sigma
_{\downarrow})]$ is a parametrizes the properties of the bulk ferromagnet
\cite{yt-stiffness}. When the spin-flip in F is negligible, \textit{i.e}.,
$d_{F}\ll \lambda_{sd}^{F}$, then $\mu_{s}^{F}|_{x=0}=s_{z}|_{x=0}\cos \theta$
and consequently the longitudinal spin current vanishes. In the present limit,
$\omega \tau_{sf}^{N}\gg1$, the time-averaged pumping current Eq.
(\ref{eq:totalcurrent}) reads $I_{s,z}^{(p)}={\hbar \omega \text{Re}%
g^{\uparrow \downarrow}\sin^{2}\theta}/{4\pi}$ and the spin accumulation in N
at distance $l_{\omega}$ near the interface becomes
\begin{equation}
s_{z}=\frac{\hbar \omega \sin^{2}\theta}{\eta_{N}(\omega)+\sin^{2}\theta
+\frac{(1-p_{\omega}^{2})\eta_{F}^{\uparrow \downarrow}}{1-p_{\omega}^{2}+\eta
_{F}}\cos^{2}\theta}\label{eq:sc}%
\end{equation}
where we have introduced the reduction factors for N and F:
\begin{equation}
\eta_{N}(\omega)=\frac{g_{N}}{g_{\omega}^{\uparrow \downarrow}}\tanh \frac
{d_{N}}{\lambda_{sd}^{N}},\quad \eta_{F}^{\uparrow \downarrow}=\frac{g_{F}%
}{g_{\omega}^{\uparrow \downarrow}}\tanh \frac{d_{F}}{\lambda_{sd}^{F}}~,
\end{equation}
where $g_{N}=h\nu_{\mathrm{dos}}AD_{N}/\lambda_{sd}^{N}$ and $\eta
_{F}=g_{\omega}^{\uparrow \downarrow}\eta_{F}^{\uparrow \downarrow}/g_{\omega}$.
With weak spin-flip in F, \textit{i.e.}, $d_{F}\ll \lambda_{sd}^{F}$, $\eta
_{F}^{\uparrow \downarrow}\approx0$ and Eq. (\ref{eq:sc}) reduces to
$s_{z}={\hbar \omega \sin^{2}\theta}/(\eta_{N}(\omega)+\sin^{2}\theta)$
\cite{brataas-sb}. Increasing the spin flip in F or the ratio $d_{F}%
/\lambda_{sd}^{F}$, the factor $\eta_{F}^{\uparrow \downarrow}$ gets larger and
the spin accumulation signal decreases accordingly. More interesting is the
chemical potential bias $\Delta \mu_{0}=\mu_{0}^{F}-\mu_{0}^{N}$ \ that builds
up across the interface, for which we find
\begin{equation}
\Delta \mu_{0}=\frac{\hbar \omega p_{\omega}\left(  \eta_{F}/2\right)  \sin
^{2}\theta \cos \theta}{\alpha_{F}\left(  \eta_{N}(\omega)+\sin^{2}%
\theta \right)  +(1-p_{\omega}^{2})\eta_{F}^{\uparrow \downarrow}\cos^{2}\theta
}~.\label{eq:fndrop}%
\end{equation}
where $\alpha_{F}=1-p_{\omega}^{2}+\eta_{F}.$ We now estimate the magnitude of
$s_{z}$ and $\Delta \mu_{0}$ for the typical systems Py$|$Al \cite{jedema2001}.
In Al the spin diffusion length is $\lambda_{sd}^{N}=500~%
\operatorname{nm}%
$, the spin-flip time $\tau_{sf}^{N}=100~%
\operatorname{ps}%
$ (at low temperature) and the density of states of Al is $\nu_{\mathrm{dos}%
}=1.5\times10^{47}~%
\operatorname{J}%
^{-1}~%
\operatorname{m}%
^{-3}$. The mixing conductance of the Py$|$Al interface in a diffuse
environment can be estimated as twice the Sharvin conductance of Al
\cite{bauer2003prb} to be $\text{Re}g_{\uparrow \downarrow}/A\approx
20\times10^{19}~\text{m}^{-2}$. The bare contact polarization is taken as
$p=0.4$ . The spin-flip length in Py is very short, around $\lambda_{sf}%
^{F}=5~\text{nm }$\cite{bass} and $(\sigma_{\uparrow}+\sigma_{\downarrow
})/\sigma_{\uparrow}\sigma_{\downarrow}$ is about $6.36\times10^{-7}~%
\operatorname{\Omega }%
\operatorname{m}%
\text{ }$\cite{fert1999mmm}. Assuming a magnetization precession cone of
$\theta=5^{\circ}$, the voltage $\Delta \mu_{0}/e$ of Py$|$Al interface as a
function of the FMR frequency is plotted in Fig. \ref{fig:pyalvol-omega}. The
induced spin accumulation in the normal metal and the voltages across the
interface as a function of $d_{F}$ are plotted in Fig. \ref{fig:pyal}. The
voltage bias across the interface, for given bulk properties of the normal
metal, is seen to saturate at large spin-flip scatterings on the F side
$d_{F}\gg \lambda_{sd}^{F}$. Spin-flip in the normal metal is detrimental to
both spin accumulation and voltage generation. On the other hand, a
transparency of the contact reduced from the Sharvin value increases the
polarization $p_{\omega}$ up to its bare interface value and\ with it the
voltage signal (up to a maximum value governed by the reduction factor
$\eta_{N}$ that wins in the limit of very low transparancy).

The angle dependence of the voltage across the interface is plotted in the
inset of Fig. \ref{fig:pyalvol-omega} in the limit of large spin flip in F
$d_{F}\gg \lambda_{sd}^{F}$. When $d_{N}\ll \lambda_{sd}^{N}$ (but still
$d_{N}\gg l_{\omega}$) we obtain the maximum value:
\begin{equation}
\Delta \mu_{0}=\frac{\hbar \omega p_{\omega}\left(  g_{F}/2g_{\omega}\right)
\sin^{2}\theta \cos \theta}{\alpha_{F}\sin^{2}\theta+(1-p_{\omega}^{2})g_{F}%
\cos^{2}\theta/g_{\omega}^{\uparrow \downarrow}} \label{eq:volmax}%
\end{equation}
given $\alpha_{F}\rightarrow1-p_{\omega}^{2}+g_{F}/g_{\omega}$. At small angle
of the magnetization precession $\theta$
\begin{equation}
\Delta \mu_{0}\overset{\theta \rightarrow0}{=}\frac{p_{\omega}g_{\omega
}^{\uparrow \downarrow}\theta^{2}}{2(1-p_{\omega}^{2})g_{\omega}}\hbar \omega~.
\end{equation}
In the opposite limit, $d_{N}\gg \lambda_{sd}^{N}$ (but $\lambda_{sd}^{N}\gg
l_{\omega}$) the voltage drop becomes
\begin{equation}
\Delta \mu_{0}=\frac{\hbar \omega p_{\omega}\left(  g_{F}/2g_{\omega}\right)
\sin^{2}\theta \cos \theta}{\alpha_{F}\left(  g_{N}/g_{\omega}^{\uparrow
\downarrow}+\sin^{2}\theta \right)  +(1-p_{\omega}^{2})g_{F}\cos^{2}%
\theta/g_{\omega}^{\uparrow \downarrow}}~, \label{eq:volmin}%
\end{equation}
which in the limit of small angle reduces to $\Delta \mu_{0}\rightarrow
{p_{\omega}g_{\omega}^{\uparrow \downarrow}\theta^{2}\hbar \omega}%
/{[2(1+g_{N}/g_{F})(1-p_{\omega}^{2})g_{\omega}+2g_{N}]}$. In both limits at
small precession angles, the voltages are proportional to $\theta^{2}$,
\textit{i.e.}, increases linearly with power intensity of the AC field. Eqs.
(\ref{eq:volmax}) and (\ref{eq:volmin}) as function of FMR frequency are
depicted in Fig. \ref{fig:pyalvol-omega} as solid and dashed lines.

In contrast to Berger \cite{berger1999}, who predicted voltage generation in
spin valves, \textit{viz}. that dynamics of one ferromagnet causes a voltage
when analyzed by a second ferromagnet through a normal metal spacer, we
consider here a simple bilayer. The single ferromagnetic layer here serves
simultaneously as a source and detector of the spin accumulation in the normal
metal layer. The presence of spin-flip scattering that allows the back-flow of
a parallel spin current is essential, and permalloy is ideal for this purpose.
The voltage bias under FMR conditions can be measured simply by separate
electrical contacts to the F and N layers. It can be detected even on a single
ferromagnetic film with normal metal contacts \cite{Azevedo2005}, provided
that the two contacts are not equivalent.

We can also study the FMR generated bias in a controlled way in the N$_{1}|$F
$|$N$_{2}$ trilayers in which the F layer is sandwiched by two normal metal
layers. The magnetization of the ferromagnet again precesses around the
$z$-axis. The thicknesses of $\text{N}1$, $\text{F}$ and $\text{N}2$ in the
transport direction are $d_{N1}$, $d_{F}$ and $d_{N2},$ respectively. The spin
diffusion length in normal metal node \textit{i} is $\lambda_{i}$. With weak
spin flip in the sandwiched ferromagnetic layer, $d_{F}\ll \lambda_{sd}^{F}$,
the spin accumulation of F at both interfaces are the same. We find that the
values of $\mu_{s}^{F}$ near the interfaces are mixtures of the interface
values of the spin accumulations in the normal metals. In other words, the two
normal metals talk to each other through F by the backflow and
the generated voltages across the interfaces are different given different contacts.
In the opposite
limit with massive spin flip in F, $d_{F}\gg \lambda_{sd}^{F}$, the strong spin
flip scattering eventually separates the spin accumulation in the two normal
metal nodes such that the \textquotedblleft exchange\textquotedblright%
\ between the two normal metals is suppressed. We then recover Eq.
(\ref{eq:saf}).

According to Eq. (\ref{eq:fndrop}) the voltage drops across the interfaces,
$\Delta{\mu}_{0}^{(1)}\equiv \mu_{0}^{F}-\mu_{0}^{N1}$ and $\Delta{\mu}%
_{0}^{(2)}\equiv \mu_{0}^{F}-\mu_{0}^{N2}$ are different for different
spin-diffusion lengths in the normal metals ($\lambda_{i}$) or different
conductances ($\operatorname{Re}g^{\uparrow \downarrow}$). For example, taking
identical normal metals but different contacts, \textit{e.g}., a clean and a
dirty one, $\Delta{\mu}_{0}^{(1)}$ and $\Delta{\mu}_{0}^{(2)}$ will be
different due to different spin-mixing conductances.

\begin{figure}[ptb]
\includegraphics[scale=0.43]{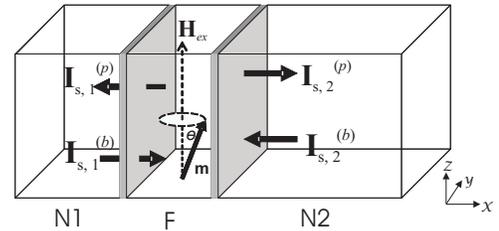}
\caption{The N$_{1}|$F $|$N$_{2}$ system in which the sandwiched F layer
precesses around the $z$-axis under FMR condition. The origin of the $x$-axis
is located at the $\text{F}|\text{N}_{\text{2}}$ interface.}%
\label{fig:trilayer}%
\end{figure}

In conclusion, we report a unified description for spin pumping in F$|$N
structure and analyze the spin accumulation in the normal metal induced by a
spin-pumping current. We predict generation of a DC voltage over a single
F$|$N junction. The Py$|$Al system should be an ideal candidate to
electrically detect magnetization dynamics in this way. An experimental test
of our predictions is in progress \cite{Costache}.

\begin{acknowledgments}
We thank Yu. Nazarov and A. Kovalev for discussions. This work is supported by
NanoNed, FOM and the Research Council of Norway through grant no. 162742/V00.
\end{acknowledgments}


\begin{thebibliography}{9999999999999999999999999999999999999999999999999999999999999999999999999}                        %
\ifx \csname natexlab\endcsname \relax


\fi
\expandafter \ifx \csname bibnamefont\endcsname \relax


\fi
\expandafter \ifx \csname bibfnamefont\endcsname \relax


\fi
\expandafter \ifx \csname citenamefont\endcsname \relax


\fi
\expandafter \ifx \csname url\endcsname \relax


\fi
\expandafter \ifx \csname urlprefix\endcsname \relax


\fi
\providecommand{\bibinfo}[2]{#2} \providecommand{\eprint}[2][]{\url{#2}}

\bibitem[Tserkovnyak et~al.(2002{a})Tserkovnyak, Brataas, and Bauer]%
{tserkovnyak-pumping}%
\bibinfo{author}{\bibfnamefont{Y.}~\bibnamefont{Tserkovnyak}},
\bibinfo{author}{\bibfnamefont{A.}~\bibnamefont{Brataas}}, and
\bibinfo{author}{\bibfnamefont{G.~E.~W.}
\bibnamefont{Bauer}}, \bibinfo{journal}{Phys. Rev. Lett}
\textbf{\bibinfo{volume}{88}}, \bibinfo{pages}{117601} (\bibinfo{year}{2002}{\natexlab{a}}).

\bibitem[Tserkovnyak et~al.(2005)Tserkovnyak, Brataas, Bauer, and
Halperin]{tserkovnyak-pumping-rmp}%
\bibinfo{author}{\bibfnamefont{Y.}~\bibnamefont{Tserkovnyak}},
\bibinfo{author}{\bibfnamefont{A.}~\bibnamefont{Brataas}},
\bibinfo{author}{\bibfnamefont{G.~E.~W.}
\bibnamefont{Bauer}}, and
\bibinfo{author}{\bibfnamefont{B.}~\bibnamefont{Halperin}},
\bibinfo{journal}{Rev. Mod. Phys.} \textbf{\bibinfo{volume}{77}},
\bibinfo{pages}{1375} (\bibinfo{year}{2005}).

\bibitem[Brataas et~al.(2002)Brataas, Tserkovnyak, Bauer, and Halperin]%
{brataas-sb}\bibinfo{author}{\bibfnamefont{A.}~\bibnamefont{Brataas}},
\bibinfo{author}{\bibfnamefont{Y.}~\bibnamefont{Tserkovnyak}},
\bibinfo{author}{\bibfnamefont{G.~E.~W.} \bibnamefont{Bauer}}, and
\bibinfo{author}{\bibfnamefont{B.~I.}
\bibnamefont{Halperin}}, \bibinfo{journal}{Phys. Rev. B}
\textbf{\bibinfo{volume}{66}}, \bibinfo{pages}{060404} (\bibinfo{year}{2002}).

\bibitem[Watts et~al.(2006)Watts, Grollier, van~der Wal, and van
Wees]{Watts2005}\bibinfo{author}{\bibfnamefont{S.~M.} \bibnamefont{Watts}},
\bibinfo{author}{\bibfnamefont{J.}~\bibnamefont{Grollier}},
\bibinfo{author}{\bibfnamefont{C.~H.} \bibnamefont{van~der Wal}}, and
\bibinfo{author}{\bibfnamefont{B.~J.} \bibnamefont{van
Wees}}, \bibinfo{journal}{Phys. Rev. Lett.} \textbf{\bibinfo{volume}{96}},
\bibinfo{pages}{077201} (\bibinfo{year}{2006}).

\bibitem[Berger(1999)]{berger1999}%
\bibinfo{author}{\bibfnamefont{L.}~\bibnamefont{Berger}},
\bibinfo{journal}{Phys. Rev. B} \textbf{\bibinfo{volume}{59}},
\bibinfo{pages}{11465} (\bibinfo{year}{1999}).

\bibitem {brouwer}J. N. Kupferschmidt, S. Adam, and P. W. Brouwer, cond-mat/0607145.

\bibitem[Azevedo et~al.(2005)Azevedo, Lei, Rodriguez-Suarez, Oliveira, and
Rezende]{Azevedo2005}%
\bibinfo{author}{\bibfnamefont{A.}~\bibnamefont{Azevedo}},
\bibinfo{author}{\bibfnamefont{L.~H.~V.}
\bibnamefont{Leo?=}}, \bibinfo{author}{\bibfnamefont{R.~L.}
\bibnamefont{Rodriguez-Suarez}}, \bibinfo{author}{\bibfnamefont{A.~B.}
\bibnamefont{Oliveira}}, and \bibinfo{author}{\bibfnamefont{S.~M.}
\bibnamefont{Rezende}}, \bibinfo{journal}{J. Appl. Phys.}
\textbf{\bibinfo{volume}{97}}, \bibinfo{pages}{10C715} (\bibinfo{year}{2005});
E. Saitoh, M. Ueda, M. Miyajima, and G. Tatara, Appl. Phys. Lett. \textbf{88},
182509 (2006).

\bibitem {Costache}M.V. Costache, M. Sladkov, C.H. van der Wal, and B.J. van
Wees, unpublished.

\bibitem[Brataas et~al.(2000)Brataas, Nazarov, and Bauer]{brataas-ct}%
\bibinfo{author}{\bibfnamefont{A.}~\bibnamefont{Brataas}},
\bibinfo{author}{\bibfnamefont{Y.~V.} \bibnamefont{Nazarov}}, and
\bibinfo{author}{\bibfnamefont{G.~E.~W.}
\bibnamefont{Bauer}}, \bibinfo{journal}{Phys. Rev. Lett}
\textbf{\bibinfo{volume}{84}}, \bibinfo{pages}{2481} (\bibinfo{year}{2000});
\bibinfo{journal}{Eur. Phys. J. B} \textbf{\bibinfo{volume}{22}},
\bibinfo{pages}{99} (\bibinfo{year}{2001}).

\bibitem[Xia et~al.(2002)Xia, Kelly, Bauer, Brataas, and Turek]{kelly}%
\bibinfo{author}{\bibfnamefont{K.}~\bibnamefont{Xia}},
\bibinfo{author}{\bibfnamefont{P.~J.} \bibnamefont{Kelly}},
\bibinfo{author}{\bibfnamefont{G.~E.~W.} \bibnamefont{Bauer}},
\bibinfo{author}{\bibfnamefont{A.}~\bibnamefont{Brataas}}, and
\bibinfo{author}{\bibfnamefont{I.}~\bibnamefont{Turek}},
\bibinfo{journal}{Phys. Rev. B} \textbf{\bibinfo{volume}{65}},
\bibinfo{pages}{220401} (\bibinfo{year}{2002});
\bibinfo{author}{\bibfnamefont{A.}~\bibnamefont{Brataas}},
\bibinfo{author}{\bibfnamefont{G.~E.~W.} \bibnamefont{Bauer}}, and
\bibinfo{author}{\bibfnamefont{P.~J.}~\bibnamefont{Kelly}},
\bibinfo{journal}{Phys. Rep.} \textbf{\bibinfo{volume}{427}},
\bibinfo{pages}{157} (\bibinfo{year}{2006}).

\bibitem[Johnson and Silsbee(1988)]{johnson-silsbee}%
\bibinfo{author}{\bibfnamefont{M.}~\bibnamefont{Johnson}} and
\bibinfo{author}{\bibfnamefont{R.~H.} \bibnamefont{Silsbee}},
\bibinfo{journal}{Phys. Rev. B} \textbf{\bibinfo{volume}{37}},
\bibinfo{pages}{5312} (\bibinfo{year}{1988}).

\bibitem[Wang and Bauer(2006)]{wangbloch}%
\bibinfo{author}{\bibfnamefont{X.}~\bibnamefont{Wang}},
\bibinfo{author}{\bibfnamefont{G.~E.~W.} \bibnamefont{Bauer}},
\bibinfo{author}{\bibfnamefont{A.}~\bibnamefont{Brataas}}, and
\bibinfo{author}{\bibfnamefont{Y.}~\bibnamefont{Tserkovnyak}}
\bibinfo{journal}{unpublished} (\bibinfo{year}{2006}).

\bibitem[Tserkovnyak et~al.(2002{b})Tserkovnyak, Brataas, and Bauer]%
{yt-stiffness}\bibinfo{author}{\bibfnamefont{Y.}~\bibnamefont{Tserkovnyak}},
\bibinfo{author}{\bibfnamefont{A.}~\bibnamefont{Brataas}}, and
\bibinfo{author}{\bibfnamefont{G.~E.~W.} \bibnamefont{Bauer}},
\bibinfo{journal}{Phys. Rev. B} \textbf{\bibinfo{volume}{67}},
\bibinfo{pages}{140404} (\bibinfo{year}{2003}{\natexlab{b}}).

\bibitem {jedema2001}\bibinfo{author}{\bibfnamefont{F.~J.}
\bibnamefont{Jedema}}, \bibinfo{author}{\bibfnamefont{H.~B.}
\bibnamefont{Heersche}}, \bibinfo{author}{\bibfnamefont{A.~T.}
\bibnamefont{Filip}}, \bibinfo{author}{\bibfnamefont{J.~J.~A.}
\bibnamefont{Baselmans}},and \bibinfo{author}{\bibfnamefont{B.~J.}
\bibnamefont{van Wees}}, \bibinfo{journal}{Nature}
\textbf{\bibinfo{volume}{416}}, \bibinfo{pages}{713} (\bibinfo{year}{2002});
\bibinfo{author}{\bibfnamefont{M.} \bibnamefont{Zaffalon}} and
\bibinfo{author}{\bibfnamefont{B.~J.} \bibnamefont{van Wees}},
\bibinfo{journal}{Phys. Rev. Lett.} \textbf{\bibinfo{volume}{91}},
\bibinfo{pages}{186601} (\bibinfo{year}{2003}).

\bibitem[Bass and Pratt(1999)]{bass}%
\bibinfo{author}{\bibfnamefont{J.}~\bibnamefont{Bass}} and
\bibinfo{author}{\bibfnamefont{W.~P.}
\bibnamefont{Pratt}}, \bibinfo{journal}{J. Magn. Magn. Mater.}
\textbf{\bibinfo{volume}{200}}, \bibinfo{pages}{274} (\bibinfo{year}{1999}).

\bibitem[Bauer et~al.(2003)Bauer, Tserkovnyak, Huertas-Hernando, and
Brataas]{bauer2003prb}%
\bibinfo{author}{\bibfnamefont{G.~E.~W.} \bibnamefont{Bauer}},
\bibinfo{author}{\bibfnamefont{Y.}~\bibnamefont{Tserkovnyak}},
\bibinfo{author}{\bibfnamefont{D.}~\bibnamefont{Huertas-Hernando}}, and
\bibinfo{author}{\bibfnamefont{A.}~\bibnamefont{Brataas}},
\bibinfo{journal}{Phys. Rev. B} \textbf{\bibinfo{volume}{67}},
\bibinfo{pages}{094421} (\bibinfo{year}{2003}).

\bibitem[Fert and Piraux(1999)]{fert1999mmm}%
\bibinfo{author}{\bibfnamefont{A.}~\bibnamefont{Fert}} and
\bibinfo{author}{\bibfnamefont{L.}~\bibnamefont{Piraux}},
\bibinfo{journal}{J. Magn. Magn.
Mater.} \textbf{\bibinfo{volume}{200}}, \bibinfo{pages}{338} (\bibinfo{year}{1999}).
\end{thebibliography}
\end{document}